\documentclass[preprint,prd,aps]{revtex4}
\usepackage{graphicx}

\begin{document}

\title{Classical Scale of Quantum Gravity}

\author{Kirill~A.~Kazakov }

\affiliation{Department of Theoretical Physics, Physics Faculty,
Moscow State University, $119899$, Moscow, Russian Federation \\
E-mail: $kirill@theor.phys.msu.ru$}

\begin{abstract}
Characteristic length scale of the post-Newtonian corrections to
the gravitational field of a body is given by its gravitational
radius $r_{\rm g}.$ The role of this scale in quantum domain is
discussed in the context of the low-energy effective theory. The
question of whether quantum gravity effects appear already at
$r_{\rm g}$ leads to the question of correspondence between
classical and quantum theories, which in turn can be unambiguously
resolved considering the issue of general covariance. The
$O(\hbar^0)$ loop contributions turn out to violate the principle
of general covariance, thus revealing their essentially quantum
nature. The violation is $O(1/N),$ where $N$ is the number of
particles in the body. This leads naturally to a macroscopic
formulation of the correspondence principle.
\end{abstract}

\maketitle

Since the early days of quantum field theory it was realized that
pursuing general quantization program in the case of gravity would
result in a theory which is very difficult to test directly,
because of the extremely small value of the characteristic quantum
length scale
$$l_{\rm P} = \sqrt{\frac{G\hbar}{c^3}}$$
that can be built from the three relevant fundamental constants of
Nature -- the Newton gravitational constant $G,$ the speed of
light $c,$ and the Planck constant $\hbar.$ Playing the role of a
coupling constant, $l_{\rm P}$ is also the root of ultraviolet
pathology in quantum gravity none of which models has yet
succeeded in reconciliation of renormalizability with unitarity
and causality. It is important, on the other hand, that smallness
of $l_{\rm P}$ fully justifies application of methods of the
effective field theory to the case of gravitational interaction.
The model independence of this approach \cite{donoghue} implies
that the low-energy properties of quantum gravity are completely
determined by the lowest-order Einstein theory whatever the
ultimate theory be. It is thus an excellent theoretic laboratory
for investigation of synthesis of quantum theory and gravitation.

In combining the characteristic parameter with dimension of length
above, one does not take into account other dimensional parameters
which can enter the theory, such as masses~($m$) of matter field
quanta. This is certainly legitimate as far as spacetime is
quantized on its own, since the quantity $l^2_{\rm P}$ is the only
parameter appearing in the quantum theory of pure gravitational
field. Inclusion of matter, however, brings in another parameter
with dimension of length, namely the gravitational radius
$$r_{\rm g} = \frac{2 G m}{c^2}\,.$$
This parameter appears, of course, already in classical theory. In
this connection, an important question arises as to whether
$r_{\rm g}$ has an independent meaning in quantum domain,
representing a scale of specifically quantum effects. This
question may seem strange at first sight, as $r_{\rm g}$ does not
contain the Planck constant $\hbar,$ an inalienable attribute of
quantum theory. However, well known is the fact that gravitational
radiative corrections do contain pieces independent of $\hbar.$ In
the framework of the effective theory, they appear as a power
series in $r_{\rm g}/r,$ just like post-Newtonian corrections in
classical general relativity. This fact was first clearly stated
by Iwasaki \cite{iwasaki}. The reason for the appearance of
$\hbar^0$ terms through the loop contributions is that in the case
of gravitational interaction, the mass and ``kinetic'' terms in a
matter Lagrangian determine not only the properties of matter
quanta propagation, but also their couplings. Thus the mass term
of, e.g., scalar field Lagrangian generates the vertices
proportional to
$$\left(\frac{m c}{\hbar}\right)^2$$
containing inverse powers of $\hbar.$ Naively, one expects these
be cancelled by $\hbar$'s coming from the propagators when
combining an amplitude. One should remember, however, that such
counting of powers of $\hbar$ in Feynman diagrams is a bad
helpmate in the presence of massless particles. Virtual
propagation of self-interacting gravitons results in a {\it root}
non-analyticity of the massive particle form factors at zero
momentum transfer ($p$). For instance, the low-energy expansion of
the diagram in Fig.~1 begins with terms proportional to
$\sqrt{-p^2}\,,$ rather than integer powers of $p^2.$ It is this
singularity which is responsible for the appearance of $r_{\rm
g}/r$-terms. The question we ask is of what nature, classical or
quantum, these pieces are.

\begin{figure}
\includegraphics{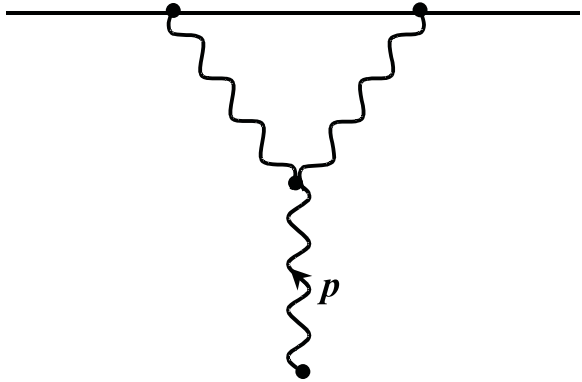}
\caption{One-loop diagram contributing to the first post-Newtonian
correction. Wavy lines represent gravitons, full lines matter
quanta. $p$ denotes momentum transfer.}
\end{figure}

Evidently, one can answer this question only after one established
a limiting procedure of transition from quantum to classical
theory. Thus, this is actually the question of {\it
correspondence} in quantum gravity.

In classical Einstein theory, gravitational field is completely
described by the metric tensor $g_{\mu\nu}\,.$ To establish a
correspondence between classical and quantum theories, one has to
find a quantum field quantity to be traced back to $g_{\mu\nu}\,.$
For this purpose, one could try to use the scattering matrix to
define a potential of particle interaction, and then compare it
with the corresponding classical quantity (the
Einstein-Infeld-Hoffman potential, in the lower orders). In fact,
this is the way followed by Iwasaki in \cite{iwasaki}.
Unfortunately, this method turned out to be highly ambiguous
\cite{hiida1,hiida2,hiida3}. There are infinitely many potentials
which lead to the same S-matrix, and one has to invent certain
{\it ad-hoc} prescriptions at each order of the post-Newtonian
expansion to achieve an agreement with classical theory. Clearly,
no valuable correspondence can be established in this way, which
could help us to elucidate nature of the post-Newtonian radiative
corrections.

There is, however, a much more direct approach to this problem,
based on calculation of the expectation value $\langle
g_{\mu\nu}\rangle \,.$ Diagrammatically, $\langle
g_{\mu\nu}\rangle$ is represented by the sum of all diagrams
having one external graviton line and an arbitrary number of
external matter lines. For instance, contribution of the order
$G^2$ is given by the sum of the tree diagrams shown in Fig.~2,
and the one-loop diagrams like that in Fig.~1. Suppose that the
matter producing gravitational field satisfies the usual quantum
mechanical quasi-classical conditions, {\it e.g.,} consider
sufficiently heavy particles. Then the quasi-classical conditions
for the gravitational field can be inferred from the requirement
that $\langle g_{\mu\nu}\rangle$ coincides with the corresponding
classical solution of the Einstein equations. Practically, the
most decisive way of looking for these conditions is to compare
{\it transformation properties} of the quantities involved under
deformations of the reference frame, thus avoiding explicit
calculation of the expectation values. The latter point of view
takes advantage of the fact that the transformation law of
classical solutions is known in advance. Hence, we have to check
whether $\langle g_{\mu\nu}\rangle$ transforms covariantly with
respect to transitions between different reference frames. In
other words, we have to consider the question of {\it general
covariance in quantum gravity.} This question might also seem
strange, since the principle of general covariance is what the
whole theory is based upon. One should remember, however, that in
quantum domain, this principle is a quite formal operator relation
expressing degeneracy of the quantum action. Transformation
properties of {\it observable} quantities is what we are
interested in instead.

\begin{figure}
\includegraphics{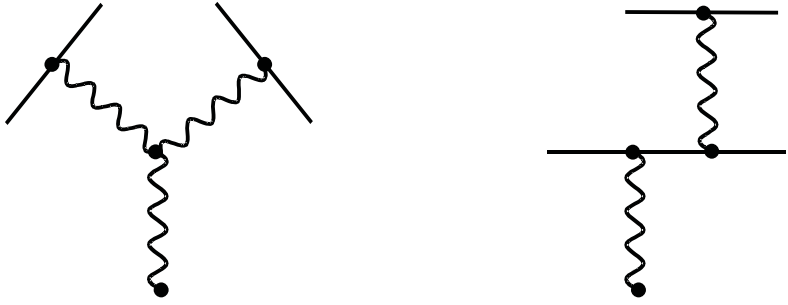}
\caption{$G^2$-order tree contribution to the effective
gravitational field.}
\end{figure}

A choice of the reference frame is equivalent to imposition of an
appropriate set of gauge conditions on the metric field.  As far
as the tree contribution to $\langle g_{\mu\nu}\rangle$ is
considered, dependence on the choice of gauge can be easily
determined in a quite general form using the anti-canonical
formalism \cite{tyutin}. The result is that gauge variations
induce spacetime diffeomorphisms, just like in classical theory.
This is as it should be, since the effective action of quantum
theory coincides with the initial classical action at the tree
level. Investigation of the post-Newtonian contributions coming
from radiative corrections is a much more difficult task because
formal manipulations using standard techniques do not give a
definite answer. An explicit calculation reveals the following
remarkable fact: gravitational radiative corrections to $\langle
g_{\mu\nu}\rangle$ transform non-covariantly under transitions
between different reference frames \cite{kazakov1}. Specifically,
let the gravitational field be produced by a scalar particle with
mass $m,$ and the reference frame be fixed by the following
conditions depending on a parameter $\varrho$
$$\eta^{\mu\nu}\partial_{\mu}g_{\nu\gamma} -
\left(\frac{\varrho - 1}{\varrho -
2}\right)\eta^{\mu\nu}\partial_{\gamma}g_{\mu\nu} = 0\,.$$
Violation of general covariance is most conveniently expressed in
terms of classically invariant quantities, e.g., the scalar
curvature $R.$ Consider deformations of the reference frame,
induced by variations of $\varrho.$ Under such a deformation, the
scalar curvature measured at a given point of the reference frame
acquires a non-zero variation resulting from the one-loop diagram
in Fig.~1 \cite{kazakov1}: $$\delta R = \frac{G^2 m^2}{c^4r^4}(1 -
2\varrho)\delta\varrho\,.$$

Thus, despite their independence of the Planck constant, the
post-Newtonian loop contributions turn out to be of a purely
quantum nature.

We are now in a position to ask for conditions to be imposed on a
system in order to allow classical consideration of its
gravitational field. Such a condition providing vanishing of the
$\hbar^0$ loop contributions can easily be found out by examining
their dependence on the number of particles in the system. Let us
consider a body with mass $M,$ consisting of a large number $N =
M/m$ of elementary particles with mass $m.$ Then it is readily
seen that the $n$-loop contribution to the effective gravitational
field of the body turns out to be suppressed by a factor $1/N^n$
in comparison with the tree contribution. For instance, at the
first post-Newtonian order, the tree diagrams in Fig.~2 are {\it
bilinear} in the energy-momentum tensor $T^{\mu\nu}$ of the
particles, and therefore proportional to $(m \cdot N) \cdot
(m\cdot N) = M^2.$ On the other hand, the post-Newtonian
contribution of the diagram in Fig.~1 is proportional to $m^2
\cdot N = M^2/N,$ since it has only two external matter lines.

Thus, we are led to the following formulation of the
correspondence principle in quantum gravity: {\it the effective
gravitational field produced by a macroscopic body of mass $M$
consisting of $N$ particles turns into corresponding classical
solution of the Einstein equations in the limit $N \to \infty$}
\cite{kazakov2}. In particular, the principle of general
covariance is to be considered as approximate, valid only for the
description of macroscopic phenomena.

The quantum gravity effects characterized by the scale $r_{\rm g}$
are normally highly suppressed. For the solar gravitational field,
their relative value is $m_{\rm proton}/M_{\odot} \approx
10^{-57}.$ However, they are the larger the more gravitating body
resembles an elementary particle, and can become noticeable for a
sufficiently massive compact body. Black hole physics is likely
the most promising place to search for manifestations of the
classical scale in quantum gravity \cite{kazakov3}.

{}

\end{document}